\documentclass[aps,twocolumn,pra,superscriptaddress,showpacs,tightenlines]{revtex4}
\usepackage{amssymb}
\usepackage{amsmath}
\usepackage{graphicx}
\usepackage{epsfig}
\usepackage{txfonts}
\usepackage{subfigure}
\usepackage{amsfonts}
\usepackage{CJK}

\begin{document}

\begin{CJK*}{GBK}{song}
\title{Amplification of quantum discord between two uncoupled qubits \\in a common environment by phase decoherence}
\author{Ji-Bing Yuan}
\affiliation{Key Laboratory of Low-Dimensional Quantum Structures
and Quantum Control of Ministry of Education, Hunan Normal
University, Changsha 410081, China and Department of Physics,
Hunan Normal University, Changsha 410081, China}
\author{Jie-Qiao Liao}
\affiliation{Key Laboratory of Low-Dimensional Quantum Structures
and Quantum Control of Ministry of Education, Hunan Normal
University, Changsha 410081, China and Department of Physics, Hunan
Normal University, Changsha 410081, China}
\affiliation{Institute of
Theoretical Physics, Chinese Academy of Sciences, Beijing 100190,
China}
\author{Le-Man Kuang\footnote{Author to whom any correspondence should be
addressed. }\footnote{ Email: lmkuang@hunnu.edu.cn}}
\affiliation{Key Laboratory of Low-Dimensional Quantum Structures
and Quantum Control of Ministry of Education, Hunan Normal
University, Changsha 410081, China and Department of Physics,
Hunan Normal University, Changsha 410081, China}
\date{\today}

\begin{abstract}
We study analytically the dynamic behaviors of quantum correlation
measured by quantum discord between two uncoupled qubits, which
are immersed in a common Ohmic environment. We show that the
quantum discord of the two noninteracting qubits can be greatly
amplified or protected for certain initially prepared $X$-type
states in the time evolution. Especially, it is found that there
does exist the stable amplification of the quantum discord for the
case of two identical qubits, and the quantum discord can be
protected for the case of two different qubits with a large
detuning. It is also indicated that in general there does exist a
sudden change of the quantum discord in the time evolution at a
critic time point $t_c$, and the discord amplification and
protection may occur only in the time interval $0<t\leq t_c$ for
certain $X$-type states. This sheds new light on the creation and
protection of quantum correlation.
\end{abstract}

\maketitle \narrowtext
\end{CJK*}

\section{\label{Sec:1}Introduction}

It is well known that the total correlation in a bipartite quantum
system can be measured by quantum mutual
information~\cite{Groisman2005,Schumacher2006}, which may be divided
into classical and quantum
parts~\cite{Groisman2005,Henderson2003,Oppenheim2002,D.
Yang2008,dlzhou2008, Ollivier2001,Kaszlikowski2008,Piani2008}. The
quantum part is called quantum discord which is originally
introduced by Olliver and Zurek~\cite{Ollivier2001}. Recently, it
has been aware of the fact that quantum discord is a more general
concept to measure quantum correlation than quantum entanglement
since there is a nonzero quantum discord in some separable mixed
states~\cite{Ollivier2001}. In fact, quantum discord is a different
type of quantum correlation than entanglement, and it can be
considered as a more universal resource than quantum entanglement in
some sense. As shown in Refs.~\cite{Datta2008,Lanyon2008}, although
there is no quantum entanglement, quantum discord can also be
responsible for the quantum computational efficiency of
deterministic quantum computation with one pure
qubit~\cite{Knill1998}. In addition, much recent attention has been
paid to many relative topics of quantum
discord~\cite{Dillenschneider2008,Sarandy2009,Cui2010,Rodriguez-Rosario2008,Rigolin2010,Wang2010,Chen2010,Alber2010,Luo2008,Lu2010},
such as quantum discord of open quantum systems.

We know that any realistic quantum systems interact inevitably with
their surrounding environments, which introduce quantum noise into
the systems. As a result, the quantum systems will lose their energy
(dissipation) and/or coherence (dephasing). Thus it is of
fundamental importance to know the influence of the environment on
quantum correlation. In several recent
papers~\cite{Werlang2009,Maziero2009,Wang2009,Maziero2010,Piilo2010},
the quantum correlation dynamics in open quantum systems have been
studied. It was shown that the quantum correlation measured by
quantum discord is more resistant against the environment than
quantum entanglement~\cite{Werlang2009}. For a certain class of
states under Markovian dynamics, the quantum entanglement can
disappear within a finite time, a phenomenon called by entanglement
sudden death~\cite{Yu2004}, which has been widely investigated in
recent
years~\cite{Yu2004,Bellomo2007,Choi2007,opez2008,Qasimi2008,Maniscalco2008}.
Differently, quantum discord only vanishes asymptotically at
infinite time~\cite{Werlang2009,Ferraro2010}. Moreover, for some
special initial states, quantum correlation in a bipartite quantum
system will not be affected by the decoherence environment during an
initial time interval~\cite{Piilo2010}.

Based on the above mentioned two facts that quantum discord is a
useful resource for quantum information processing, and that quantum
systems couple inevitably with their environment. Naturally there
are two interesting questions: (i) if the environment can enhance
the quantum discord of the systems to realize the discord
amplification? (ii) if we can obtain a stable quantum discord
induced by the environment? With these questions, in this paper, we
study the dynamics of quantum discord between two non-interacting
qubits immersed in a common Ohmic environment. We show that it is
possible to amplify and protect the quantum discord by the
qubit-environment interaction under some conditions. Especially,
when the two qubits are identical, we find that the phase
decoherence can induce a stable amplification of the
initially-prepared quantum discord for certain $X$-type states.

This paper is organized as follows. In Sec.~\ref{Sec:2}, we
present our physical model and its solution. In Sec.~\ref{Sec:3},
we investigate dynamical behaviors of quantum discord for the
so-called $X$-type initial states. In Sec.~\ref{Sec:4}, we study
analytically and numerically the quantum discord for certain
$X$-type initial states, and show the initially prepared discord
can be amplified or protected for $X$-type states. Finally, we
conclude this work in Sec.~\ref{Sec:5}.

\section{\label{Sec:2} physical model and solution}

\begin{figure}[tbp]
\includegraphics[bb=35 643 318 774, width=3.2 in]{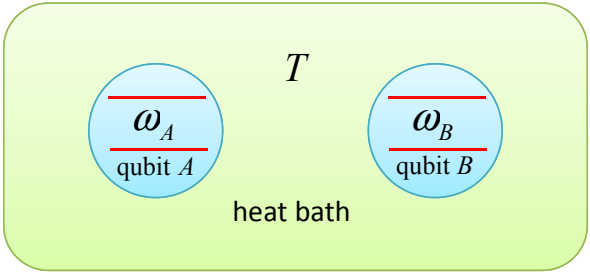}
\caption{(Color online) Schematic of our physical system: two
uncoupled qubits, of energy separations $\omega_{A}$ and
$\omega_{B}$, are immersed in a common heat bath with temperature
$T$.} \label{fig1.eps}
\end{figure}

Let us start with introducing the physical system, as shown in
Fig.~\ref{fig1.eps}, two uncoupled qubits, qubit $A$ and qubit $B$
with energy separations $\omega_{A}$ and $\omega_{B}$ respectively,
are immersed in a common environment. The Hamiltonian of the total
system including the two qubits and the environment is composed of
four parts,
\begin{equation}
\label{1}
\hat{H}_{T}=\hat{H}_{S}+\hat{H}_{E}+\hat{H}_{I}+\hat{H}_{R},
\end{equation}
where $\hat{H}_{S}$ is the Hamiltonian of the two qubits,
$\hat{H}_{E}$ is the Hamiltonian of the environment, $\hat{H}_{I}$
represents the interaction Hamiltonian between the two qubits and
the environment, and $\hat{H}_{R}$ is a renormalization term, which
is introduced originally in Ref.~\cite{Caldeira1983}.

The expression of the Hamiltonian $\hat{H}_{S}$ reads
\begin{equation}
\label{2} \ \hat{H}_{S}=\frac{1}{2}\omega _{A}\hat{\sigma}
_{A}^{z}+\frac{1}{2}\omega _{B}\hat{\sigma} _{B}^{z},
\end{equation}
where  $\hat{\sigma} _{A(B)}^{z}=\left\vert 0\right\rangle
_{A(B)}\left\langle 0\right\vert -\left\vert 1\right\rangle
_{A(B)}\left\langle 1\right\vert $ with $ \left\vert 0\right\rangle
_{A(B)}$ and $\left\vert 1\right\rangle _{A(B)}$ being the excited
and ground states of the qubit $A$ ($B$). Hereafter, we take $\hbar
=1$.

The environment of the two qubits is modelled by a heat bath with
temperature $T$, which is composed of an infinite set of harmonic
oscillators with the following Hamiltonian,
\begin{equation}
 \label{3}
 \hat{H}_{E} =\sum_{k}\omega _{k}\hat{b}_{k}^{\dag }\hat{b}_{k},
\end{equation}
where $\omega _{k}$ is the frequency of the $k$th harmonic
oscillator depicted by the usual bosonic creation and annihilation
operators $\hat{b}_{k}^{\dag }$ and $\hat{b} _{k}$, satisfying the
commutative relation $[\hat{b}_{k},\hat{b}_{k}^{\dag }]=1$.

As for the interaction Hamiltonian $\hat{H}_{I}$ between the two
qubits and the heat bath, we assume it is given by the following
expression
\begin{equation}
\label{4} \hat{H}_{I}
=\hat{H}_{S}\sum_{k}g_{k}(\hat{b}_{k}^{\dag}+\hat{b}_{k}),
\end{equation}
where $g_{k}$ is the coupling constant between the system and the
$k$th harmonic oscillator of the heat bath. It is obvious that
$\hat{H}_{I}$ commutes with the Hamiltonian $\hat{H}_{S}$. Therefore
there is no energy dissipation in this system. Note that this form
of coupling has been used to study quantum decoherence in
Bose-Einstein condensation and trapped ion by Kuang and
coworkers~\cite{Kuang1999}.

The renormalization term has the following form
\begin{equation}
\label{5} \hat{H}_{R}
=(\hat{H}_{S})^{2}\sum_{k}\frac{g_{k}^{2}}{\omega _{k}^{2}}.
\end{equation}

According to Ref.~\cite{Kuang1999}, Hamiltonian~(\ref{1}) can be
exactly solved by making use of the unitary transformation
\begin{equation}
\label{6} \hat{U}=\exp
\left[\hat{H}_{S}\sum_{k}g_{k}(\hat{b}_{k}^{\dag }-\hat{b}_{k})
\right].
\end{equation}

Corresponding to Hamiltonian~(\ref{1}), the total density operator
of the system plus the heat bath can be expressed as
\begin{equation}
\label{7}
\hat{\rho}_{T}(t)=e^{-i\hat{H}_{S}t}\hat{U}^{-1}e^{-it\sum_{k}\omega
_{k} \hat{b}_{k}^{\dag }\hat{b}_{k}}\hat{U}\hat{\rho}_{T}(0)\hat{U}
^{-1}e^{it\sum_{k}\omega _{k}\hat{b}_{k}^{\dag
}\hat{b}_{k}}\hat{U}e^{i\hat{ H }_{s}t},
\end{equation}
where $\hat{\rho}_{T}(0)$ is the initial total density operator.

We assume the system and the heat bath are initially uncorrelated
with the density operator $\hat{\rho}_{T}(0)$ =$\hat{\rho}(0)\otimes
\hat{\rho}_{R}(0)$, where $\hat{\rho}(0)$ is the initial density
operator of the system, and $\hat{\rho}_{R}(0)$ is the density
operator of the heat bath, which is assumed to be
$\hat{\rho}_{R}=\prod_{k}{\otimes}\hat{\rho}_{k}(0),$ where
$\hat{\rho}_{k}(0)$ is the density operator of the $k$th harmonic
oscillator in thermal equilibrium of temperature $T$. We can obtain
the reduced density operator of the system, denoted by $\hat{\rho}
(t)=\textrm{Tr}_{R}[\hat{\rho} _{T}(t)]$. Its matrix elements in the
eigen representation of $\hat{H} _{s}$ (with the four eigenstates
$\left\vert 00\right\rangle,\left\vert 01\right\rangle ,\left\vert
10\right\rangle ,\left\vert 11\right\rangle$) can be written as
\begin{eqnarray}
\label{8} \rho _{(l'_{A},l'_{B})(l_{A},l_{B})}(t) &=&\rho
_{(l'_{A},l'_{B})(l_{A},l_{B})}(0)R_{(l'_{A},l'_{B})(l_{A},l_{B})}(t)
\notag \\
&&\times e^{-i[E(l'_{A},l'_{B})-E(l_{A},l_{B})]t}, \label{Pt}
\end{eqnarray}
where $E(l_{A},l_{B})$ $(E(l'_{A},l'_{B}))$ is the eigenvalue of
the operator $\hat{H}_{S}$ with the corresponding eigenstate
$\vert l_{A},l_{B}\rangle$ $(\vert l'_{A},l'_{B}\rangle)$, the
expression of $E(l_{A},l_{B})$ is
$E(l_{A},l_{B})=[(-1)^{l_{A}}\omega_{A}+(-1)^{l_{B}}\omega_{B}]/2$.
The quantity $R_{(l'_{A},l'_{B})(l_{A},l_{B})}(t)$ is a
reservoir-dependent part given by
\begin{eqnarray}
\label{9}
R_{(l'_{A},l'_{B})(l_{A},l_{B})}(t) &=&e^{-i[E^{2}(l'_{A},l'_{B})-E^{2}(l_{A},l_{B})]Q_{1}(t)}  \notag \\
&&\times
e^{-[E(l'_{A},l'_{B})-E(l_{A},l_{B})]^{2}Q_{2}(t)}.\label{Rmnt}
\end{eqnarray}
The two reservoir-dependent functions $Q_{1}(t)$ and $Q_{2}(t)$ in
the above Eq.~(\ref{9}) are given by
\begin{subequations}
\label{10}
\begin{align}
Q_{1}(t)&=\int_{0}^{\infty }d\omega J(\omega )\frac{g^{2}(\omega
)}{\omega
^{2}}\sin (\omega t), \\
Q_{2}(t)&=2\int_{0}^{\infty }d\omega J(\omega )\frac{g^{2}(\omega
)}{\omega
^{2}}\sin ^{2}\left(\frac{\omega t}{2}\right)\coth \left(\frac{\beta \omega }{%
2}\right).
\end{align}
\end{subequations}
Here we have taken the continuum limit of the reservoir modes
$\sum_{k}\rightarrow $ $\int_{0}^{\infty }d\omega J(\omega )$, where
$ J(\omega )$ is the spectral density of the reservoir, $g(\omega )$
is the corresponding continuum expression for $g_{k}$, and $\beta
=1/T$ (with the Boltzmann constant $k_{B}=1$).

\section{\label{Sec:3}dynamics of quantum discord for two-qubit $X$-type states}

In this section, we investigate dynamics of quantum discord for the
two non-interacting qubits in the phase decoherence environment for
the three-parameter two-qubit $X$-type states. Quantum
discord~\cite{Ollivier2001} is defined as the difference between the
total correlation and the classical correlation with the following
expression
\begin{equation}
\label{11} \mathcal {D}\left( \hat{\rho}\right) =\mathcal{I}\left(
\hat{\rho}_{A}:\hat{\rho}_{B}\right)
-\mathcal{C}\left(\hat{\rho}\right).
\end{equation}
Here the total correlation in a bipartite quantum state
$\hat{\rho}$ is measured by quantum mutual information given by
\begin{eqnarray}
\label{12} \mathcal{I}\left( \hat{\rho} _{A}:\hat{\rho} _{B}\right)
=S\left( \hat{\rho} _{A}\right) +S\left( \hat{\rho} _{B}\right)
-S\left( \hat{\rho}\right),
\end{eqnarray}
where  $S\left( \hat{\rho} \right) =-\textrm{Tr}(\hat{\rho} \log
\hat{\rho})$ is the von Neumann entropy,
$\hat{\rho}_{A}=\textrm{Tr}_{B}(\hat{\rho})$ and
$\hat{\rho}_{B}=\textrm{Tr}_{A}(\hat{\rho})$ are  the reduced
density operators for subsystems $A$ and $B$, respectively. And the
classical correlation between the two subsystems $A$ and $B$ can be
defined as
\begin{eqnarray}
\label{13} C(\hat{\rho})&=&\max_{\{\hat{P}_{k}\}}\left[ S(\hat{\rho}
_{A})-\sum_{k}p_{k}S(\hat{\rho}_{A}^{(k)})\right]\nonumber\\
&=&S(\hat{\rho} _{A})-\min_{\{\hat{P}_{k}\}}\left[
\sum_{k}p_{k}S(\hat{\rho}_{A}^{(k)})\right]. \label{clacor}
\end{eqnarray}
Here $\{\hat{P}_{k}\}$ is a set of projects performed locally on the
subsystem $B$, and $\hat{\rho}
_{A}^{(k)}=\frac{1}{p_{k}}\textrm{Tr}_{B}\left[ \left(
\hat{I}_{A}\otimes \hat{P}_{k}\right) \hat{\rho} (\hat{I}_{A}\otimes
\hat{P}_{k})\right]$ is the state of the subsystem $A$ conditioned
on the measurement of the outcome labelled by $k$ ,where
$p_{k}=\textrm{Tr}_{AB}[(\hat{I}_{A}\otimes
\hat{P}_{k})\hat{\rho}(\hat{I}_{A}\otimes \hat{P}_{k})]$ denotes the
probability relating to the outcome $k$, and $\hat{I}_{A}$ denotes
the identity operator for the subsystem $A$.

In terms of the relation given in Eq.~(\ref{8}), we can study the
quantum discord dynamic properties of the two qubits. We assume the
two qubits are initially prepared in a class of state with maximally
mixed marginals ($\hat{\rho} _{A(B)}=\hat{I}_{A(B)}/2$) described by
the three-parameter $X$-type density matrix
\begin{eqnarray}
\label{14}
\hat{\rho}(0)&=&\frac{1}{4}\left(\hat{I}_{AB}+\underset{i=1}{\overset{3}{\sum
}}c_{i}\hat{\sigma}_{A}^{i}\otimes \hat{\sigma} _{B}^{i}\right)
\nonumber \\
&=&\frac{1}{4}\left(
\begin{array}{cccc}
1+c_{3} & 0 & 0 & c_1-c_2 \\
0 & 1-c_{3} & c_1+c_2 & 0 \\
0 & c_1+c_2 & 1-c_{3} & 0 \\
c_1-c_2 & 0 & 0 & 1+c_{3}
\end{array}
\right), \label{AB}
\end{eqnarray}
where $\hat{I}_{AB}$ is the identity operator in the Hilbert space
of the two qubits, $\hat{\sigma} _{A}^{i}$ and $\hat{\sigma}
_{B}^{i}$ ($i=1,2,3$ mean $x,y,z$ correspondingly) are the Pauli
operators of qubit $A$ and qubit $B$, and $c_{i}$ ($0\leq \left\vert
c_{i}\right\vert \leq 1$) are real numbers satisfying the unit trace
and positivity conditions of the density operator $\hat{\rho}$. The
density operator $\hat{\rho}$ includes the Werner states and the
Bell states as two special cases. Under the decoherence environment,
the evolution of density operator $\hat{\rho}(t)$ initially prepared
in Eq.~(\ref{14}) can be obtained according to Eq.~(\ref{8}). Its
explicit form at time $t$ is
\begin{equation}
\label{15} \hat{\rho}(t)=\frac{1}{4}\left(
\begin{array}{cccc}
1+c_{3} & 0 & 0 & \mu e^{i\Delta _{1}} \\
0 & 1-c_{3} & \nu e^{i\Delta _{2}} & 0 \\
0 & \nu e^{-i\Delta _{2}} & 1-c_{3} & 0 \\
\mu e^{-i\Delta _{1}} & 0 & 0 & 1+c_{3}
\end{array}
\right),
\end{equation}
where we have introduced the following parameters
\begin{eqnarray}
\label{16}
\Delta _{1}&=&(\omega _{A}+\omega _{B})t,\hspace{0.9cm} \Delta _{2}=(\omega_{A}-\omega _{B})t, \nonumber \\
\mu(t)&=&(c_{1}-c_{2})\gamma _{1}(t),\hspace{0.5cm} \nu(t) =(c_{1}+c_{2})\gamma _{2}(t),\nonumber \\
\gamma_{1}(t)&=&e^{-(\omega_{A}+\omega_{B})^{2}Q_{2}(t)},\hspace{0.3cm}\gamma_{2}(t)=e^{-(\omega_{A}-\omega_{B})^{2}Q_{2}(t)}.\label{r12}
\end{eqnarray}

As shown in Eq.~(\ref{16}), the decay parameters $\gamma_{1}$ and
$\gamma_{2}$ depend on the frequencies $\omega_A$ and $\omega_B$ of
the two qubits and the reservoir-dependent function $Q_{2}$. For
convenience, we define a detuning parameter of the two qubits as
\begin{equation}
\label{17} r=\frac{\omega _{A}}{\omega _{B}},
\end{equation}
which indicates that the detuning parameter $r=1$ for two identical
qubits due to $\omega _{A}=\omega _{B}$, and  $r\neq 1$ for two
different qubits due to $\omega _{A}\neq\omega _{B}$. In other
words, the two qubits are resonant (non-resonant) when the detuning
parameter $r=1$ ($r\neq 1$). Then the two decay parameters $\gamma
_{1}$ and $\gamma _{2}$ can be connected with each other through the
detuning parameter with the following simple expression
\begin{equation}
\label{18} \gamma _{2}=\gamma
_{1}^{\left(\frac{r-1}{r+1}\right)^{2}}.
\end{equation}

When the two qubits are identical, i.e., $r=1$, from Eq.~(\ref{16})
we have $\Delta_2=0$,  $\gamma _{2}=1$, and $\nu =(c_{1}+c_{2})$. In
this case, equation~(\ref{15}) indicates that there does exist a
decoherence-free subspace with two basis states $|1,0\rangle_{AB}$
and $|0,1\rangle_{AB}$ for the $X$-type initial states under present
consideration.

In order to obtain the mutual information of state $\hat{\rho}(t)$
given in Eq.~(\ref{15}), we first calculate the four eigenvalues of
$\hat{\rho}(t)$,
\begin{eqnarray}
\label{19} \lambda _{1,2} &=&\frac{1}{4}(1+c_{3}\mp \mu
),\hspace{0.5 cm} \lambda _{3,4} =\frac{1}{4}(1-c_{3}\mp \nu ).
\label{eigval}
\end{eqnarray}
Then the mutual information reads
\begin{equation}
\label{20} \mathcal{I}\left( \hat{\rho}_{A}:\hat{\rho}_{B}\right) \
=2+{\sum_{i=1}^{4}}\lambda _{i}\log\lambda _{i}.
\end{equation}
Note that here we have used
$S(\hat{\rho}_{A}(t))=S(\hat{\rho}_{B}(t))=1$, since the two reduced
density matrixes $\hat{\rho}_{A}(t)$ and $\hat{\rho}_{B}(t)$ are
maximally mixed, that is $\hat{\rho}_{A}(t)=\hat{I}_{A}/2$ and
$\hat{\rho}_{B}(t)=\hat{I}_{B}/2$.

For calculation of the amount for the classical correlation
$C(\hat{\rho})$ defined in Eq.~(\ref{13}), we propose the complete
set of orthogonal projectors $\{\hat{P}_{k}=|\theta _{k}\rangle
\langle \theta _{k}|, k=\Vert ,\perp \}$ for a local measurement
performed on the subsystem $B$, where the two projectors are defined
in terms of the following two orthogonal states
\begin{subequations}
\label{21}
\begin{align}
\left\vert \theta _{\parallel }\right\rangle &=\cos \theta
\left\vert 0\right\rangle +e^{i\phi }\sin \theta \left\vert
1\right\rangle,\\
\left\vert \theta _{\perp }\right\rangle &=e^{-i\phi }\sin \theta
\left\vert 0\right\rangle -\cos \theta \left\vert 1\right\rangle,
\end{align}
\end{subequations}
with $0\leq \theta \leq \pi /2$ and $0\leq \phi \leq 2\pi$. After
the two project measurements with
$p_{\shortparallel}=p_{\bot}=1/2$ , the reduced density matrices
of subsystem $A$ read
\begin{subequations}
\label{22}
\begin{align}
\hat{\rho} _{A}^{\shortparallel }=\frac{1}{4}\left(
\begin{array}{cc}
2(1-c_{3}\cos (2\theta) ) & \epsilon \sin (2\theta)  \\
\epsilon ^{\ast }\sin (2\theta)  & 2(1+c_{3}\cos (2\theta) )
\end{array}\right),\\
\hat{\rho} _{A}^{\bot }=\frac{1}{4}\left(
\begin{array}{cc}
2(1+c_{3}\cos (2\theta) ) & -\epsilon \sin (2\theta)  \\
-\epsilon ^{\ast }\sin (2\theta)  & 2(1-c_{3}\cos (2\theta))
\end{array}
\right),
\end{align}
\end{subequations}
where we have introduced the parameter $\epsilon=\mu e^{i(\Delta
_{1}-\phi )}+\nu e^{i(\Delta _{2}+\phi )}$. According to
Eq.~(\ref{22}), it is straightforward to obtain the eigenvalues of
the reduced density matrix $\hat{\rho} _{A}^{(k)}$ as follows:
\begin{eqnarray}
\label{23} \zeta _{1,2}^{(k)} &=&\frac{1}{2}(1\pm
\Lambda),\label{eiva}
\end{eqnarray}
where we have defined $\Lambda$ as
\begin{eqnarray}
\label{24} \Lambda&=&\left\{ c_{3}^{2}\cos ^{2}(2\theta
)+\frac{1}{4}\left[\mu ^{2}+\nu ^{2}+2\mu \nu \cos (\Delta
_{2}-\Delta
_{1}+2\phi )\right]\right.   \notag \\
&&\left. \times\sin ^{2}(2\theta )\right\} ^{\frac{1}{2}}.
\end{eqnarray}
Since $\zeta _{1,2}^{(\parallel )}$ =$\zeta _{1,2}^{(\perp )}$,
then we have  $S(\hat{\rho}_{A}^{\parallel
})=S(\hat{\rho}_{A}^{\perp })=f(\Lambda)$ defined by
\begin{eqnarray}
\label{25}
f(\Lambda)=-\frac{1-\Lambda}{2}\log_{2}\left(\frac{1-\Lambda}{2}\right)-\frac{1+\Lambda}{2}\log_{2}\left(\frac{1+\Lambda}{2}\right),
\end{eqnarray}
which leads to the classical correlation
\begin{equation}
\label{26} \mathcal {C}\left(\hat{\rho}(t)\right)
=1-\underset{\theta ,\phi }{\min }\left[ f(\Lambda)\right].
\end{equation}

Since the function $ f(\Lambda)$ is a monotonically decreasing
function, therefore for getting the minimal value of
$S(\hat{\rho}_{A}^{\parallel })=f(\Lambda)$, we should choose proper
parameters $\theta$ and $\phi$ to ensure the parameter $\Lambda$
defined in Eq.~(\ref{24}) is maximal. Obviously, from Eq.~(\ref{24})
we can see that the maximal value depends on $c_{3}, \mu$, and
$\nu$. From Eq.~(\ref{24}) it is easy to get the following
inequality
\begin{eqnarray}
\label{27} \Lambda&\leq&\left[c_{3}^{2}\cos ^{2}(2\theta
)+\frac{(|\mu|+|\nu|)^{2}}{4}\sin ^{2}(2\theta )\right]
^{1/2}\nonumber\\
&\leq&\left\{
\begin{array}{c}
|c_{3}|,\hspace{1.3 cm} \textrm{for} \hspace{0.2 cm} |c_{3}|>(|\mu|+|\nu|)/2,\\
(|\mu|+|\nu|)/2,\hspace{0.3 cm} \textrm{for}
\hspace{0.2cm}|c_{3}|<(|\mu|+|\nu|)/2.
\end{array}
\right.
\end{eqnarray}

If we define $\chi(t)$ as
\begin{equation}
\label{28} \chi(t) =\max \left[\left\vert
c_{3}\right\vert,(|\mu(t)|+|\nu(t)|) /2\right]
\end{equation}
then the classical correlation can be expressed as
\begin{equation}
\label{29} \mathcal{C}(\hat{\rho}(t))
=\overset{2}{\underset{n=1}{\sum }}\frac{1+(-1)^{n}\chi }{ 2}\log
_{2}\left[ 1+(-1)^{n}\chi \right].
\end{equation}

Therefore, the quantum discord can be written as
\begin{eqnarray}
\label{30} \mathcal{D}\left(\hat{\rho}(t)\right) \
&=&2+{\sum_{i=1}^{4}}\lambda _{i}\log _{2}\lambda
_{i}-\mathcal{C}(\hat{\rho}(t)),
\end{eqnarray}
where the amount of the classical correlation
$\mathcal{C}(\hat{\rho}(t))$ is given by Eq.~(\ref{29}). In
principle, we have obtained the dynamics of the quantum discord
according the above expression given in Eq.~(\ref{30}), provided
that we know the initial condition of the system. In what follows we
will study dynamic properties of the quantum discord for some
initial states in detail.

\section{\label{Sec:4}quantum discord amplification}

In the section, we would like to find the possibility of quantum
discord enhancement induced by phase decoherence in the dynamic
evolution of the $X$-type quantum states. From Eq.~(\ref{16}) we can
see that both $\mu(t)$ and $\nu(t)$ are two decaying functions with
respect to the evolution time $t$ due to the reservoir function
$Q_2(t)$ being a increase function indicated below, hence there may
exist a critic time $t_c$. At this time, the following equation may
be satisfied
\begin{equation}
\label{31} \frac{|\mu(t_c)|+|\nu(t_c)|}{2}=\left\vert
c_{3}\right\vert,
\end{equation}
which is the equation to determine $t_c$. Obviously, the critic
time  $t_c$ depends on initial-state and reservoir parameters.

From Eqs.~(\ref{16}) and (\ref{28}) we can obtain
\begin{eqnarray}
\label{32} \chi (t) &=\left\{
\begin{array}{c}
|c_{3}|,\hspace{2.1 cm} \textrm{for}  \hspace{0.2 cm} t>t_c,\\
(|\mu(t)|+|\nu(t)|)/2,\hspace{0.3 cm} \textrm{for} \hspace{0.2 cm}
t<t_c,
\end{array}
\right.
\end{eqnarray}
which indicates that the classical correlation expressed by
Eq.~(\ref{29}) exhibits a sudden change at the critic time $t_c$
determined by Eq.~(\ref{31}). The classical correlation decays
monotonically from the initial time to the critic time $t_{c}$, and
then keeps constant after the critic time $t_{c}$. Then making use
of Eqs.~(29-32) we can obtain expressions of the quantum discord in
different regimes of the time evolution
\begin{subequations}
\label{33}
\begin{align}
\label{33a} \mathcal{D}\left( \hat{\rho}(t)\right)
=&2+{\sum_{i=1}^{4}}\lambda _{i}\log _{2}\lambda _{i}-\frac{1}{2}\sum_{n=1}^{2}\Lambda'_n\log _{2}\Lambda'_n, \hspace{0.2 cm} t<t_{c}, \\
\label{33b}\mathcal{D}\left(\hat{\rho}(t)\right)
=&2+{\sum_{i=1}^{4}}\lambda _{i}\log _{2}\lambda _{i}
-\frac{1}{2}\sum_{n=1}^{2}\Lambda_n\log _{2}\Lambda_n, \hspace{0.2
cm} t\geq t_{c},
\end{align}
\end{subequations}
where $\Lambda_n$ and $\Lambda'_n$ are given by
\begin{equation}
\label{34} \Lambda_n=1+(-1)^{n}\left\vert c_{3}\right\vert,
\hspace{0.5cm} \Lambda'_n=1+ (-1)^n\frac{|\mu|+|\nu|}{2}.
\end{equation}

From Eqs.~(\ref{33a}) and (\ref{33b}) we can see that there does
exist a critic time at which the quantum discord evolution exhibits
a sudden change. Therefore, dynamic evolution of the quantum discord
may have different behaviors in the time interval $0\leq t \leq t_c$
and after the critic time. In what follows, we shall investigate
dynamic characteristics of the quantum discord in the time evolution
for two identical and different qubits, respectively.

\subsection{The case of two identical qubits}

For the sake of simplicity, we first consider the case of two
identical qubits and assume that the three initial-state parameters
obey $c_2=0$ and $c_1>c_3>0$. In this case, we have
$\omega_A=\omega_B=\Omega$, $r=1$ and $\gamma_2=1$. From
Eq.~(\ref{31}) we find that the critic time $t_c$ satisfies the
equation
\begin{equation}
\label{35} Q_2(t_c)=\frac{1}{4\Omega^2}\ln
\left(\frac{2c_3}{c_1}-1\right),
\end{equation}
where $Q_2(t)$ is the reservoir function defined in Eq.~(\ref{10})
which is determined by the spectral density of the environment
$J(\omega)$. We consider the Ohmic reservoir \cite{Chakravarty1984}
with the spectral density $J(\omega)$ given by
\begin{equation}
\label{36}
 J(\omega )=\frac{\eta \omega
}{g^{2}\left( \omega \right) }\exp\left(-\frac{\omega }{\omega
_{c}}\right),
\end{equation}
where $\omega _{c}$ is the high-frequency cutoff, and $\eta $ is a
positive characteristic parameter of the reservoir. With this
choice, at low temperature the function  $Q_{2}(t)$ is given by the
following expression
\begin{equation}
\label{37}
 Q_{2}(t) =\eta \left\{ \frac{1}{2}\ln \left[ 1+(\omega
_{c}t)^{2}\right] +\ln \left[ \frac{\beta }{\pi t}\sinh
\left(\frac{\pi t}{\beta }\right) \right] \right\} .
\end{equation}

The the critic time $t_c$ can be obtained from Eqs.~(\ref{35})
and~(\ref{37}). However, it should be pointed out that there does
exist the critic time only for certain values of the three
initial-state parameters $(c_1, c_2, c_3)$, not for arbitrary values
of these three parameters. Take into account the positivity of the
density operator and the reservoir function given by Eq.~(\ref{37}),
we find that there exists the critic time $t_c$ for some initial
states with $0<c_1/2\leq c_3< c_1\leq 2/3$ and $c_2=0$. It is a
difficult task to get an analytical expression of the critic time
$t_c$ for an arbitrary reservoir temperature. However, at zero
temperature, from Eqs.~(\ref{35}) and (\ref{37}) it is
straightforward to get the analytic result of the critic time given
by
\begin{equation}
\label{38}
t_c=\frac{1}{\omega_c}\sqrt{\left(2c_3/c_1-1\right)^{-1/(2\eta
\Omega^2)}-1}.
\end{equation}

\begin{figure}[tbp]
\includegraphics[width=3.2 in]{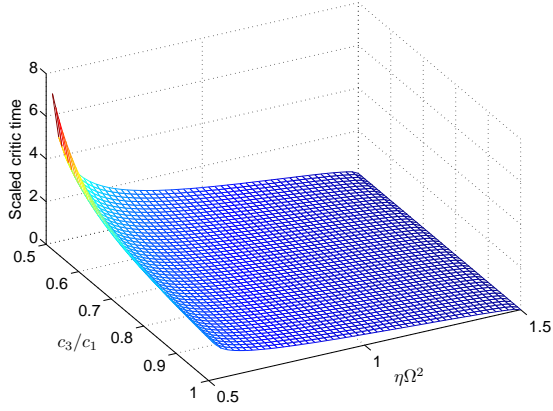}
\caption{(Color online)  Plot of the scaled critic time
$\omega_ct_c$ given in Eq.~(38) vs the qubit parameter
$\eta\Omega^{2}$ and the initial-state parameter $c_3/c_1$ at zero
temperature.} \label{fig2.eps}
\end{figure}

From Eq.~(\ref{38}) we can see that the critic time $t_c$ depends on
not only the characteristics of the two  qubits and their
environment but also the initial-state parameters. For a set of
given parameters ($c_1, c_3, \eta, \Omega$), the lower is the
cut-off frequency of the reservoir, the longer is the critic time.
In particular, when $2c_3=c_1$ we have $t_c=\infty$. This means that
the quantum discord will not exhibit the sudden change in the whole
process of the time evolution. In Fig.~\ref{fig2.eps} we display the
critic time with respect to the parameters $\eta\Omega^{2}$ and
$c_3/c_1$ at zero temperature. In the following we show that at zero
temperature it is possible to realize the discord amplification in
the case of $2c_3=c_1$ for the two identical qubits.

For the two identical qubits, when $2c_3=c_1$  and $c_2=0$ the
critic time approaches the infinity. Then the quantum discord is
given by
\begin{equation}
\label{39} \mathcal{D}_{id}(t)=2+\sum_{i=1}^{4}\lambda _{i}\log
_{2}\lambda _{i} -\frac{1}{2}\sum_{n=1}^{2}\Lambda' _{n}\log
_{2}\Lambda' _{n}, \hspace{0.2 cm}0\leq t < \infty,
\end{equation}
where $\lambda _{i}$, $\Lambda' _{n}$, and $\gamma_1(t)$ are given
by
\begin{eqnarray}
\label{40}\lambda _{1,2}&=&\frac{1}{8}[2+c_1(1\mp 2\gamma_1(t))],
\hspace{0.3 cm} \lambda
_{3,4}=\frac{1}{8}[2-c_1(1\pm2)], \nonumber \\
\Lambda' _{1,2}&=&1\mp\frac{1}{2}c_1(1+\gamma_1(t)), \hspace{0.5 cm}
\gamma_1(t)=e^{-4\Omega^2Q_2(t)},
\end{eqnarray}
where the temperature-dependent reservoir function $Q_2(t)$ is given
by Eq.~(\ref{37}).

\begin{figure}[tbp]
\includegraphics[width=3.2 in]{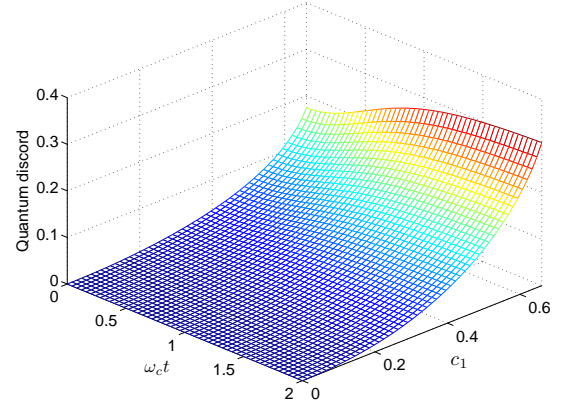}
\caption{(Color online)  Plot of the quantum discord
$\mathcal{D}_{id}(t)$ for the two identical qubits  vs the scaled
time $\omega_c t$ and the initial-state parameter $c_1$ at zero
temperature when $\eta\Omega^{2}=1$.} \label{fig3.eps}
\end{figure}

\begin{figure}[tbp]
\includegraphics[width=3.2 in]{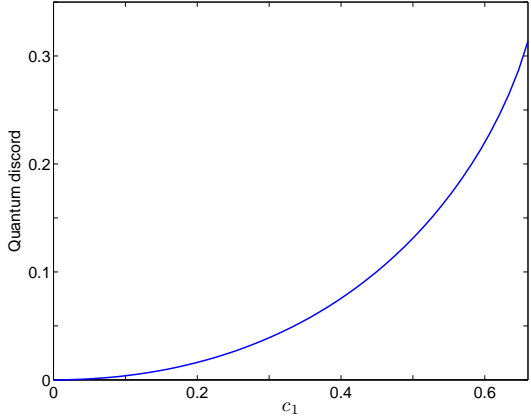}
\caption{(Color online)  Plot of the quantum discord given in Eq.
(41) with respect to the initial-state parameter $c_1$.}
\label{fig4.eps}
\end{figure}

In Fig.~\ref{fig3.eps} we have plotted the quantum discord
$\mathcal{D}_{id}(t)$ with respect to the initial-state parameter
$c_1$ in the time evolution. Figure~\ref{fig3.eps} clearly indicates
the amplification of the initial discord in the time evolution. And
the discord amplification becomes more apparent for large values of
$c_1$. In particular, when the time approaches to the infinity, we
have $\gamma_1=0$, hence the quantum discord reaches its maximal
value given by
\begin{eqnarray}
\label{41} \mathcal{D}_{id}(\infty)&=&\frac{2+c_1}{8}\log
_{2}(2+c_1)-\frac{2-c_1}{4}\log _{2}(2-c_1)\nonumber\\
&&+\frac{2-3c_1}{8}\log _{2}(2-3c_1),
\end{eqnarray}
which implies that the interaction between the reservoir and the two
identical qubits can lead to the stable amplification of the
initially prepared discord. And the stable value of the quantum
discord depends on only the initial-state parameters. In
Fig.~\ref{fig4.eps} we display the stable discord
$\mathcal{D}_{id}(\infty)$ with respect to the initial-state
parameter $c_1$. From Fig.~\ref{fig4.eps} we can see that the larger
is the initial-state parameter $c_1$, the more apparent the
amplification of the quantum discord.

In order to calculate the amplification rate of the quantum
discord, we need the quantum discord of the corresponding initial
states with the following expression
\begin{eqnarray}
\label{42} \mathcal{D}(0)&=&-1+\frac{2-c_1}{8}\log _{2}(2-c_1)+
\frac{2+c_1}{8}\log
_{2}(2+c_1)\nonumber\\
&&+\frac{2+3c_1}{8}\log _{2}(2+3c_1)+\frac{2-3c_1}{8}\log
_{2}(2-3c_1)\nonumber\\
&&-\frac{1+c_1}{2}\log _{2}(1+c_1)-\frac{1-c_1}{2}\log
_{2}(1-c_1).
\end{eqnarray}

\begin{figure}[tbp]
\includegraphics[width=3.2 in]{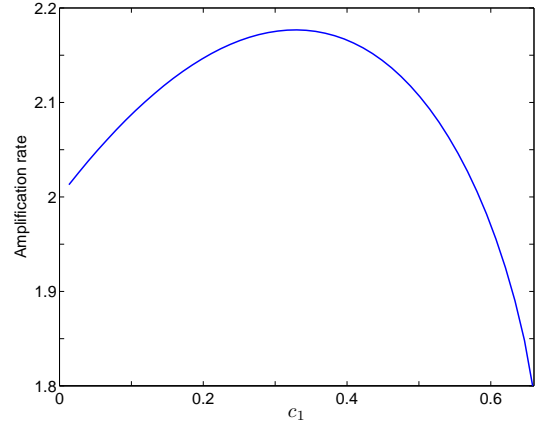}
\caption{(Color online)  Plot of the amplification rate of the
quantum discord $\Gamma=\mathcal{D}_{id}(\infty)/\mathcal{D}(0)$
with respect to the initial-state parameter $c_1$.} \label{fig5.eps}
\end{figure}

Then we define the amplification rate of the quantum discord as
$\Gamma=\mathcal{D}_{id}(\infty)/\mathcal{D}(0)$. In Fig. 5 we have
plotted the amplification rate $\Gamma$ with respect to the
initial-state parameter $c_1$. Figure~\ref{fig5.eps} indicates that
the initially-prepared discord under present consideration can be
amplified for the whole regime in which $c_1$ can takes its values,
and the amplification rate is larger in the middle regime than in
other regimes for the values of $c_1$. In particular, we can obtain
the maximal amplification rate $\Gamma_{\max}\approx 2.17$ when the
initial-state parameter $c_1\approx 3.3$.

Physically, the stable amplification phenomenon of the quantum
discord for the two resonant qubits in the common heat bath is
related to the existence of a decoherence-free subspace of the two
qubits in the time evolution. In fact, from Eq. (15) we can see
that when the two qubits are resonant, i.e., $r=1$, we have
$\nu(t)=0$, the subspace formed by $|0,1\rangle_{AB}$ and
$|0,1\rangle_{AB}$ is a decoherence-free subspace of the two
qubits in the time evolution. Therefore, the stable amplification
phenomenon of the quantum discord can be understood as a quantum
coherent effect of the two qubits in the common heat bath.

\subsection{The case of two different qubits}

Dynamic evolution of the quantum discord for two uncoupled
different qubits in a common environment can be studied through
analyzing the first derivative of the quantum discord with respect
to time. In this case, we have $r\neq 1$ due to $\omega_A\neq
\omega_B$. We now discuss the general conditions to realize the
amplification of the quantum discord by phase decoherence. Note
that the time dependence of the quantum discord given by
Eq.~(\ref{33}) can be described in terms of only one parameter
$\gamma _{1}(t)$  since the other time-dependent parameters
$\lambda _{i}$, $\Lambda'_n$, and $\gamma _{2}(t)$ can be
expressed by $\gamma _{1}(t)$ through the relations given by
Eqs.~(\ref{16}) and (\ref{17}). For the Ohmic-reservoir case under
our consideration and at low temperature, making use of
Eq.~(\ref{36}) we can find $\gamma _{1}(t)$ to be
\begin{equation}
\label{43}\gamma _{1}(t) =\left\{ \left[ 1+(\omega
_{c}t)^{2}\right]^{1/2} \frac{\beta }{\pi t} \sinh \left(\frac{\pi
t}{\beta }\right)\right\} ^{-\eta (r+1)^{2}\omega _{B}^{2}}.
\end{equation}

From Eq.~(\ref{44}), it is easy to see that the first derivative
of $\gamma _{1}$ with respect to time $t$ is always negative,
i.e.,
\begin{eqnarray}
\label{44} \frac{\partial \gamma_{1}}{\partial t}<0,
\end{eqnarray}
which means that the parameter $\gamma_{1}$ monotonically
decreases with respect to time $t$. In fact,
$\gamma_{1}=\gamma_{2}=1$ at the beginning and $\gamma
_{1}\rightarrow 0$, $\gamma _{2}\rightarrow 0$ at infinite time.

Hence, the derivative of the quantum discord with respect to
evolution time can be expressed as
\begin{equation}
\label{45} \frac{\partial\mathcal {D}\left(\hat{\rho}(t)\right)
}{\partial t}=\frac{\partial \mathcal {D}\left( \hat{\rho}(t)\right)
}{\partial \gamma _{1}}\frac{\partial \gamma _{1}}{
\partial t},
\end{equation}
which implies that when
\begin{equation}
\label{46} \frac{\partial \mathcal{D}\left(\hat{\rho}(t)\right)
}{\partial \gamma _{1}}<0,
\end{equation}
and making use of Eq.~(45) we can find
\begin{equation}
\label{47} \frac{\partial\mathcal {D}\left(\hat{\rho}(t)\right)
}{\partial t}>0,
\end{equation}
which indicates that the quantum discord is enhanced with the time
evolution. This is the condition of the amplification of the
quantum discord. In the following we will further study the
amplification condition.

In order to demonstrate the possibility of the discord
amplification, we calculate the first derivative of quantum discord
with respect to $\gamma _{1}$. Substitution of Eqs.~(\ref{18})
,~(\ref{19}), and~(\ref{34}) into Eq.~(\ref{33}), the first
derivative of $\mathcal {D}\left(\hat{\rho} (\gamma _{1})\right)$
with respect to $\gamma_{1}$ has the following forms
\begin{eqnarray}
\label{48} \frac{\partial \mathcal{D}\left( \hat{\rho}(\gamma
_{1})\right) }{\partial \gamma _{1}} &=\left\{
\begin{array}{c}
F(\gamma _{1}) + G(\gamma _{1}),\hspace{1cm} 0<t<t_{c},\\
G(\gamma _{1}), \hspace{2cm} t\geq t_{c},
\end{array}
\right.
\end{eqnarray}
where we have introduced the following three functions
\begin{eqnarray}
\label{49} F(\gamma _{1})&=& \frac{\left\vert c_{1}-c_{2}\right\vert
+\left\vert h\right\vert }{4}\log _{2}\left( \frac{2-|\mu |-|\nu
|}{2+|\mu|+|\nu |}\right), \nonumber\\
G(\gamma _{1})&=&\frac{c_{1}-c_{2}}{4}\log _{2}\left( \frac{
1+c_{3}+\mu }{1+c_{3}-\mu }\right)+ \frac{h}{4}\log _{2}\left(
\frac{1-c_{3}+\nu }{1-c_{3}-\nu }\right), \nonumber \\
h(\gamma _{1})&=&\frac{\nu}{\gamma_1}\left( \frac{r-1}{r+1}\right)
^{2}.
\end{eqnarray}
\begin{figure}[tbp]
\includegraphics[width=3.2 in]{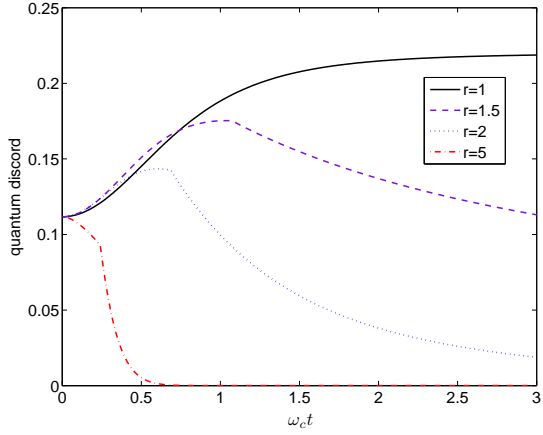}
\caption{(Color online)  Plot of the quantum discord given in
Eq.~(\ref{33}) vs the scaled time $\omega_{c}t$ at zero temperature
for different values of the detuning parameter $r$: $r=1$ (solid
black line), $r=1.5$ (dashed purple line), $r=2$ (dotted blue line),
and $r=5$ (dot-dashed red line). Other parameters are set as
$c_{1}=0.6$, $c_{2}=0$, and $c_{3}=0.3$.} \label{fig6.eps}
\end{figure}

For the initial state of $0\leq c_2<c_1<1$ and $|c_3|<1$, from
Eq.~(\ref{49}) we can see that the function $G(\gamma _{1})$ is
always positive. This implies that when $t>t_c$ the derivative of
the quantum discord with respect to the time is always negative,
i.e., $\partial \mathcal{D}\left( \hat{\rho}(\gamma
_{1})\right)/\partial t<0$. Hence, the quantum discord always
decreases monotonically when $t>t_c$. That is, it is impossible to
amplify the prepared quantum discord in the time evolution after the
critic time $t_c$.

On the other hand, from Eqs.~(\ref{48}) and (\ref{49}) we can see
that the function $F(\gamma _{1})$ is always negative due to
$(2-|\mu |-|\nu |)/(2+|\mu |+|\nu |)<1$. Then,  $\partial
\mathcal{D}\left( \hat{\rho}(\gamma _{1})\right)/\partial \gamma_1
<0$, i.e., $\partial \mathcal{D}\left( \hat{\rho}(\gamma
_{1})\right)/\partial t <0$, is possible due to the competing change
of the two functions $F(\gamma _{1})$ and $G(\gamma _{1})$ in the
time evolution when $t<t_c$. Therefore, it is possible to amplify
the initially prepared quantum discord before the critic time.

We now investigate numerically the discord amplification for certain
initially prepared $X$-type states of two different qubits. In
Fig.~\ref{fig6.eps} we display the dynamic evolution of the quantum
discord at zero temperature. Here the two qubits are initially
prepared in the $X$-type state with $c_1=0.6$, $c_2=0$, and
$c_3=0.3$, the detuning parameter $r=1, 1.5, 2$, and $5$,
respectively. The solid black line corresponds to the case of two
identical qubits discussed in the previous subsection. In this case,
the initially prepared discord can be amplified and this discord
amplification is stable with the critic time $t_c=\infty$. From Fig.
6 we can see that the detuning of the two qubits described by the
parameter $r$ can affect seriously the discord evolution. The critic
time at which the sudden change of the discord happens becomes
shorten with the increase of the detuning parameter $r$. After the
critic time, the discord begins to decay asymptotically. There
exists an amplification regime of the discord before the critic
time. The amplification regime becomes narrow with the increase of
$r$, and the amplification disappears when the detuning is large
enough.

The above discussions on the discord dynamics inspire us to ask an
interesting question: may the discord remain unchange in the time
evolution before the critic time for certain cases? In the following
we show that it is possible to maintain the discord before the
critic time. In order to see this, we consider two different qubits
with a large detuning $r \gg 1$, which are initially prepared
$X$-type states with the  three state parameters  $c_1=1$, and
$0<-c_2=c_3=c<1$. In this case, at zero temperature we find the
critic time is given by
\begin{equation}
\label{50} t_c=\frac{1}{\omega_c}\sqrt{c^{-2/(\eta
 \omega^2_A)}-1},
\end{equation}
and the quantum discord has the following expressions
\begin{eqnarray}
\mathcal{D}\left( \hat{\rho}\right)&=\left\{
\begin{array}{c}
\frac{1+c}{2}\log_2 (1+c) +\frac{1-c}{2}\log_2 (1-c),\hspace{0.2cm} t<t_{c},\\
\frac{1+\gamma_1}{2}\log_2 (1+\gamma_1)
+\frac{1-\gamma_1}{2}\log_2 (1-\gamma_1), t\geq t_{c},
\end{array}
\right.\label{51}
\end{eqnarray}
where $\gamma_1(t)$ is given by Eq.~(\ref{43}). From Eq.~(\ref{51})
we can see that before the critic time the quantum discord is
independent of the time. This indicates that the initial discord of
the $X$-type states under present consideration does not change in
the time evolution before the critic time $t_c$. Hence, the
environment does not affect the quantum discord in this time
evolution regime. In other words, the quantum discord is
decoherence-free for such $X$-type states with the three state
parameters $c_1=1$, and $0<-c_2=c_3=c<1$ in the time evolution
before the critic time.  From Eq.~(\ref{50}) we can see that the
smaller the value of the state parameter $c$ the longer the discord
can remain unchanged.

\section{\label{Sec:5} Concluding remarks}

In conclusion, we have studied the quantum discord dynamics of two
uncoupled qubits immersed in a common heat bath. This system is
depicted by an exactly solvable phase decoherence model. We have
shown that the quantum discord of two noninteracting qubits can be
amplified or protected for certain initially prepared $X$-type
states in the time evolution. Especially, it has been found that
there does exist the stable amplification of the quantum discord
for certain $X$-type states when the two qubits are identical, and
the quantum discord can be protected for the case of two different
qubits with a large detuning.  However, the stable amplification
cannot be created when the two qubits are not identical for the
same $X$-type states. The degree of the discord amplification
decreases with the increase of the detuning between the two
qubits. The amplification and protection of the quantum discord
can be understood as quantum  effects induced by the environment
since the two qubits interact neither directly , nor with a third
system, but with the common reservoir with infinitely many degrees
of freedoms. Both of them are decoherence free. The stable
amplification (protection) of the quantum discord can also be
considered as a resonant (large-detuning) effect of the two qubits
interacting with the common reservoir.

It has been indicated that in general there does exist a sudden
change of the quantum discord for the two qubits in the time
evolution at a critic time point $t_c$, and the discord
amplification and protection may occur only in the time interval
$0<t\leq t_c$ for certain $X$-type states. Generally, the critic
time depends on the initial-state parameters, characteristic
parameters of the two qubits and their environment, and the
qubit-reservoir interaction. The quantum discord will decay
asymptotically to zero after the critic time. The creation of the
stable amplification and protection of the quantum discord for the
two qubits in a heat bath sheds new light on production of quantum
states with long-living quantum discord.

\acknowledgments This work was supported by  NFRP of China Grant
No. 2007CB925204, NSF of China Grant No. 10775048, PCSIRT Grant
No. IRT0964, and ECHP Grant No. 08W012. J. B. Yuan thanks Qiong
Wang, Qin-Shou Tan, Mi Jiang, and Yu-Xia Shu for useful
discussions.

\end{document}